\documentclass{amsart}

\usepackage{epsfig,a4wide}

\newtheorem{theorem}{Theorem}
\newtheorem{proposition}{Proposition}

\newcommand{\N}{\ensuremath{\mathbb{N}^{}_0}}
\newcommand{\Z}{\ensuremath{\mathbb{Z}}}
\newcommand{\R}{\ensuremath{\mathbb{R}}}
\newcommand{\Q}{\ensuremath{\mathbb{Q}}}

\newcommand{\Sb}{\ensuremath{\mathbb{S}}}
\newcommand{\X}{\ensuremath{\mathbb{X}}}

\newcommand{\D}{\ensuremath{{\mathcal D}}}
\newcommand{\T}{\ensuremath{{\mathcal T}}}
\newcommand{\gL}{\varLambda}
\newcommand{\gG}{\varGamma}

\newcommand{\bs}{\boldsymbol}
\newcommand{\smc}{\scriptscriptstyle\square}

\begin{document}

\title[Pinwheel patterns and powder diffraction]{\textbf{Pinwheel patterns and powder diffraction}}

\author{Michael Baake}
\address{Fakult\"at f\"ur Mathematik, Universit\"at Bielefeld, Postfach
  100131, 33501 Bielefeld,  Germany}
\email{\texttt{mbaake@math.uni-bielefeld.de}, 
\texttt{dirk.frettloeh@math.uni-bielefeld.de}}
\urladdr{\texttt{http://www.math.uni-bielefeld.de/baake/},
\texttt{http://www.math.uni-bielefeld.de/baake/frettloe/}}

\author{Dirk Frettl\"oh}
 
\author{Uwe Grimm}
\address{Department of Mathematics, The Open University, Walton Hall,
  Milton Keynes MK7 6AA, UK}
\email{\texttt{u.g.grimm@open.ac.uk}}
\urladdr{\texttt{http://mcs.open.ac.uk/ugg2/}}

\begin{abstract}
Pinwheel patterns and their higher dimensional generalisations display
continuous circular or spherical symmetries in spite of being
perfectly ordered. The same symmetries show up in the corresponding
diffraction images. Interestingly, they also arise from amorphous
systems, and also from regular crystals when investigated by powder
diffraction. We present first steps and results towards a general
frame to investigate such systems, with emphasis on statistical
properties that are helpful to understand and compare the diffraction
images. We concentrate on properties that are accessible via an
alternative substitution rule for the pinwheel tiling, based on two
different prototiles. Due to striking similarities, we compare our
results with the toy model for the powder diffraction of the square
lattice.
\end{abstract}

\maketitle

\section{Pinwheel patterns}

\begin{figure} 
\begin{center}
\epsfig{width=\textwidth,file=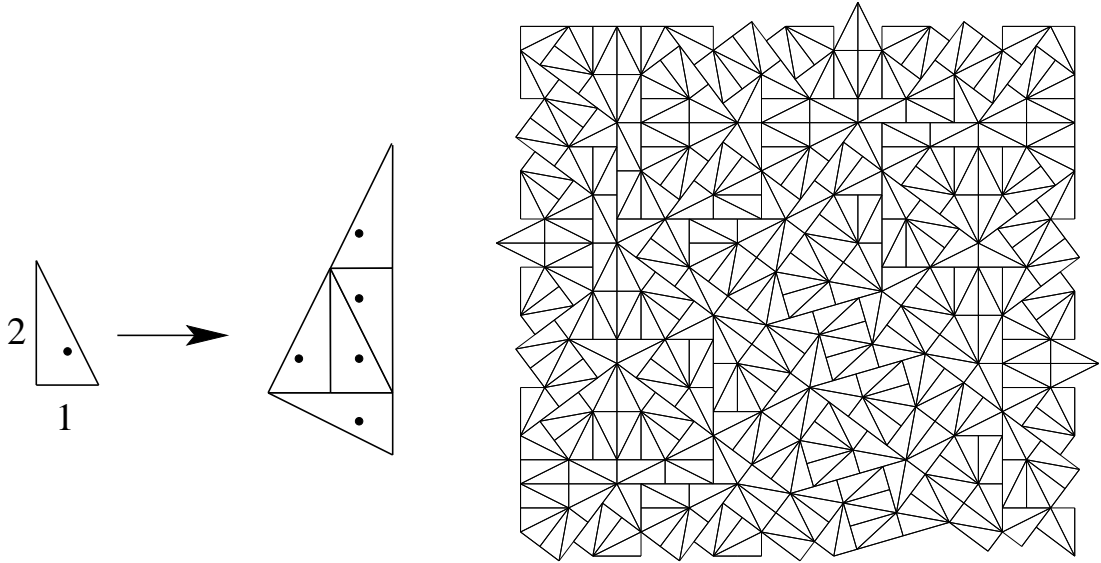}
\end{center}
\caption{The pinwheel substitution rule and a patch of the pinwheel
  tiling $\T$. The points in the left part indicate how the point set 
  $\gL=\gL_{\T}$ 
  arises from the pinwheel tiling. \label{fig:pinw}} 
\end{figure}

The Conway-Radin pinwheel tiling \cite{rad}, a variant of which is
shown in Figure~\ref{fig:pinw}, is a substitution tiling with tiles
occurring in infinitely many orientations. Consequently, it is not of
\emph{finite local complexity} (FLC) with respect to translations
alone, though it is FLC with respect to Euclidean motions. This
property distinguishes the pinwheel tiling from the majority of
substitution tilings considered in the literature. As a consequence,
its diffraction differs considerably from that of other tilings, and
despite a growing interest in such structures \cite{ors,mps,bfg,yok},
the diffraction properties have only been partially understood to
date.

\begin{figure} 
\begin{center}
\epsfig{width=\textwidth,file=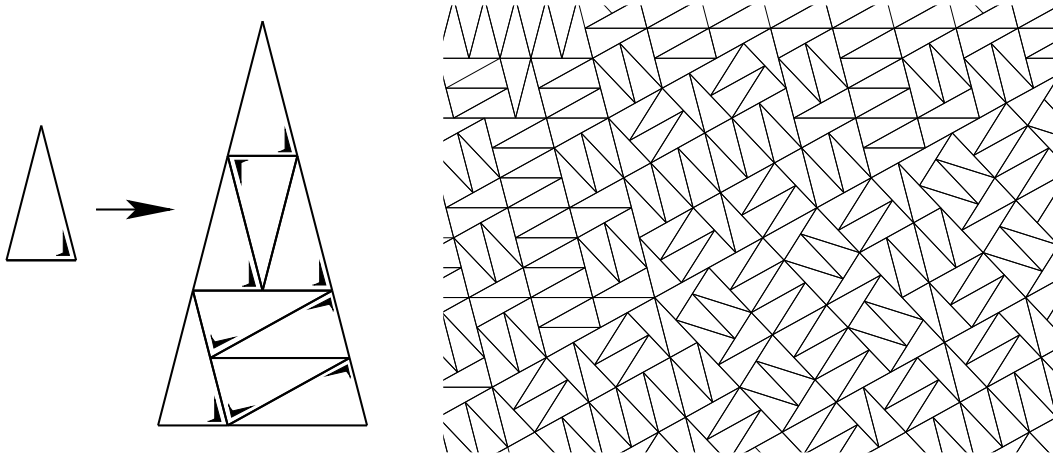}
\end{center}
\caption{Another substitution rule which generates a tiling with
  circularly symmetric autocorrelation. The decoration of the triangles
  indicates that, in contrast to the classic pinwheel, no enantiomorphic 
  pairs of triangles occur here.
 \label{fig:pinw9}} 
\end{figure}

Whereas the pinwheel tiling is the most commonly investigated example,
there are other tilings with infinitely many orientations, compare
\cite{sgen} for an entire family of generalisations.  Yet another
example is shown in Figure~\ref{fig:pinw9}. It has a single prototile,
an equilateral triangle with side lengths $1$, $2$ and $2$.  Under
substitution, the prototile is mapped to nine copies, some rotated by
an angle $\theta= \arccos(1/4)$, which is incommensurate to $\pi$
(i.e., $\theta \notin \pi \Q$). Thus, the corresponding rotation
$R_{\theta}$ is of infinite order, and the tiles occur in infinitely
many orientations in the infinite tiling.  Here and below,
$R_{\alpha}$ denotes the rotation through the angle $\alpha$ about the
origin. More examples of tilings with tiles in infinitely many
orientations can be found in \cite{fh}.

It was shown constructively in \cite{mps} that the autocorrelation
$\gamma$ of the pinwheel tiling has full circular symmetry, a result
that was implicit in previous work \cite{rad}. As a consequence, the
diffraction measure $\widehat{\gamma}$ of the pinwheel tiling shows
full circular symmetry as well. To make this concrete, we now
construct a Delone set from the tiling. Recall that a \emph{Delone
set} $\gL$ in Euclidean space is a point set which is uniformly
discrete (i.e., there is $r>0$ such that each ball of radius $r$
contains at most one point of $\gL$) and relatively dense (i.e., there
is $R>0$ such that each ball of radius $R$ contains at least one point
of $\gL$). Let $\T$ be the unique fixed point of the pinwheel
substitution of Figure~\ref{fig:pinw} that contains the triangle with
vertices $(\frac{1}{2},-\frac{1}{2}), (-\frac{1}{2},-\frac{1}{2}),
(-\frac{1}{2},\frac{3}{2}) $.  This fixed point $\T$ is the same as
the one considered in \cite{mps}. We now define the set of
\emph{control points} $\gL_{\T}$ of $\T$ to be the set of all points
${\bs u} + \frac{{\bs u} - {\bs v}}{2} + \frac{{\bs u} - {\bs w}}{4}$
such that the triangle with vertices ${\bs u},{\bs v},{\bs w}$ is in
$\T$ and $\overline{{\bs u}{\bs v}}$ is the edge of length one. This
choice of control points is indicated in Figure~\ref{fig:pinw} (left)
and is the same as in \cite{mps}.

Recall that the natural autocorrelation measure of a Delone set $\gL$ is
defined as 
\begin{equation}\label{auto-1}
\gamma  \; := \; \lim_{R \to \infty} \frac{1}{\pi R^2}
\sum_{x,y \in \gL \cap B_R } \delta_{x-y}, 
\end{equation} 
where the limit is taken in the vague topology and exists in all
examples discussed below; for details, see \cite{bl,hof,sch}. Here,
$\delta_{x}$ denotes the Dirac measure in $x$, and $B_R$ the closed
ball of radius $R$ centred at the origin. The Fourier transform
$\widehat{\gamma}$ is then the diffraction measure of $\gL$, whose
nature is often the first property to be analysed.  Since
$\widehat{\gamma}$ is a translation bounded measure on $\R^2$, it has
a unique decomposition, relative to Lebesgue measure, into three parts,
\[ 
\widehat{\gamma} \; =\;  \widehat{\gamma}^{}_{\sf pp} +
\widehat{\gamma}^{}_{\sf sc} + \widehat{\gamma}^{}_{\sf ac}, 
\] 
where the pure point part $\widehat{\gamma}^{}_{\sf pp}$ is a
countable sum of (weighted) Dirac measures, $\widehat{\gamma}^{}_{\sf
ac}$ is absolutely continuous with respect to Lebesgue measure, and
$\widehat{\gamma}^{}_{\sf sc}$ is supported on a set of Lebesgue
measure $0$, but vanishes on single points.

It was shown in \cite{mps} that the autocorrelation $\gamma^{}_{\gL}$ of
the pinwheel control points $\gL=\gL_{\T}$ satisfies 
\begin{equation} \label{eq:etarmur}
\gamma^{}_{\gL} \; =\; \delta_0 + \sum_{r \in \D \setminus \{0\}}
\eta(r) \mu_r \; =\;  \sum_{r \in \D} \eta(r) \mu_r,  
\end{equation}
where $\D$ is a discrete subset of $[0,+\infty)$, $\mu_r$ denotes the
normalised uniform distribution on the circle $r \Sb^1 =\{ x \in \R^2
\mid |x| = r \}$, and $\eta(r)$ is a positive number. Note that
$\mu^{}_0=\delta^{}_0$.  In particular, $\gamma^{}_{\gL}$ shows
perfect circular symmetry, as does the diffraction measure
$\widehat{\gamma}^{}_{\gL}$. This settles the pure point part: Since
$\gamma^{}_{\gL}$ is a translation bounded measure, and the Fourier
transform of such a measure is also translation bounded, it follows
from the circular symmetry that there are no Bragg peaks except at
$0$. Moreover, a standard argument \cite{hof} gives
\[ 
\widehat{\gamma}^{}_{\sf pp} \; = \; \bigl(\text{dens}(\gL)\bigr)^2\, 
\delta^{}_0 \; =\;  \delta^{}_0\, ,
\]
because the density $\text{dens}(\gL)$, i.e., the average number of
points of $\gL$ per unit area, is $1$. This follows from the fact
that, in our setting, each triangle has unit area and carries
precisely one control point.

Below, we give more detailed information about $\D$ and $\eta(r)$,
which is needed to shed some light on the nature of
$\widehat{\gamma}^{}_{\sf sc}$ and $\widehat{\gamma}^{}_{\sf ac}$.

\begin{proposition} \label{distset}
The pinwheel Delone set $\gL$ as defined in Figure~$\ref{fig:pinw}$ satisfies:
\begin{itemize}
\item[\rm (i)] $\gL \subset \bigcup_{n \in \Z} R_{n \theta} \Z^2$, where
  $\theta := 2 \arctan(\frac{1}{2})$.\medskip
\item[\rm (ii)] $\gL \subset \bigl\{ (\textstyle{\frac{n}{5^{k}},
    \frac{m}{5^{k}}}) \mid m,n \in \Z, \; k\in\N \bigr\}$.
\item[\rm (iii)] The distance set $\D = \D^{}_{\gL} :=\{ |x-y| \mid x, y \in
  \gL \}$ is a subset of\/  $\bigl\{ \sqrt{ \frac{p^2+q^2}{5^{\ell}}} \mid
  p,q,\ell \in \N \bigr\}$.
\end{itemize} 
\end{proposition} 

In plain words, $\gL$ is a uniformly discrete subset of a countable
union of rotated square lattices, all elements of $\gL$ have rational
coordinates, and, as a consequence, all squared distances between
points in $\gL$ are rational numbers of the form
$(p^2+q^2)/5^{\ell}$. The fact that $\gL$ is supported on such a
simple set is an interesting property of the pinwheel tiling. It is
not clear whether a similar property, for a suitable choice of control
points, can be expected for other examples, such as that of
Figure~\ref{fig:pinw9}.

\begin{figure} 
\begin{center}
\epsfig{width=\textwidth,file=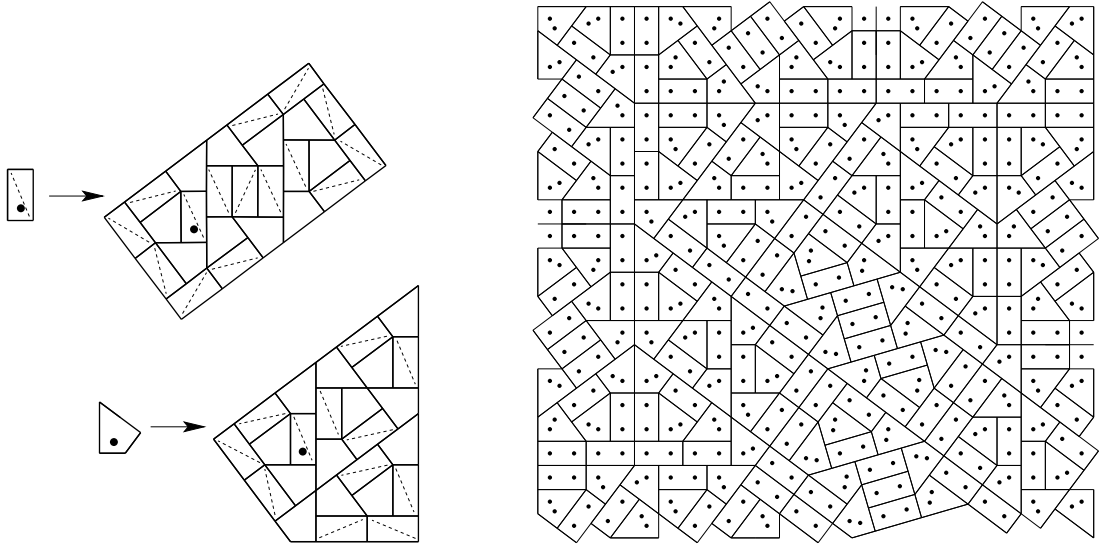}
\end{center}
\caption{The kite domino substitution rule and a patch of a kite
  domino tiling. This patch is equivalent to the pinwheel patch in
  Figure~\ref{fig:pinw}. The dots in the patch indicate how $\gL$
  arises from the tiling. \label{fig:kitedom}} 
\end{figure}

These results were obtained by means of an alternative substitution,
the \emph{kite domino} substitution shown in Figure~\ref{fig:kitedom},
which generates the same Delone set $\gL$.  The kite domino
substitution is equivalent to the pinwheel substitution in the sense
that the corresponding tilings are \emph{mutually locally derivable} (MLD) 
in the sense of \cite{bsj}, i.e., they can be obtained from each other by
local replacement rules. Moreover, the Delone set $\gL$ is MLD with 
both tilings.

Because of the strong linkage between the diffraction spectrum of a
Delone set $\gL$ and the dynamical spectrum of the associated
dynamical system $(\X(\gL),\R^d)$, we consider the \emph{hull} of $\gL$
\[ 
\X(\gL)\; = \;\overline{\R^2 + \gL}^{\,\sf LRT},
\] 
where completion is with respect to the local rubber topology (LRT),
see \cite{bl} and references therein for details.  Roughly speaking,
and restricted to the special case under consideration, this means
that $\X (\gL)$ contains all translates of $\gL$ and all Delone sets
which are locally congruent to some translate of $\gL$.

In addition to Proposition~\ref{distset}, the kite domino substitution
gives access also to the frequency of configurations in the tilings.
The \emph{frequency} of a finite set $L \subset \gL$ is defined as
\[ 
  \text{freq}(L) \; = \; \lim_{R \to \infty} \frac{1}{\pi R^2} \; 
  \text{card} \{ F 
  \subset \gL\cap B_{R} \mid F\; \text{is congruent to}\; L \}. 
\]  
Note that this definition is up to congruence of the finite sets, not
up to translation (which is not reasonable here).  The \emph{frequency
module} of $\gL$ is the $\Z$-span of $\{ \text{freq}(L) \mid L \subset
\gL \; \text{finite} \}$.

\begin{proposition}
The frequency module of $\gL$ is $\{ \frac{m}{264 \cdot 5^{\ell}} \mid
m \in \Z, \, \ell \in \N \}$. It is the same for all elements of
$\X(\gL)$.  In particular, all $\eta(r)$ in \eqref{eq:etarmur} are
rational.
\end{proposition}

Using the kite domino substitution, one can determine some frequencies
of small distances exactly. These are given below, together with some
other values (marked by an asterisk) where the frequencies are
estimated by analysing large approximants of the pinwheel tiling.
\begin{equation}\label{freqlist}
\begin{array}{l||c|c|c|c|c|c|c|c|c|c|c|c|c}
\rule[-2mm]{0mm}{7mm} r^2 & \; 0 \; & \frac{1}{5} & 1 & \; \frac{8}{5} \; &
\frac{9}{5} & \frac{49}{25} & 2 & \frac{13}{5} & \frac{81}{25} &
\frac{17}{5} & 4 & \frac{113}{25} & 5 \\ \hline 
\rule[-2mm]{0mm}{7mm} \eta(r) & 1 & \frac{5}{11} &
\frac{439}{165} & \frac{1}{2} & \frac{67}{165} & \frac{4}{165}
&  \frac{7}{2}^{\ast} & \frac{142}{165} & \frac{4}{165} &
\frac{10}{11}^{\ast} &  3^{\ast} &  \frac{8}{165}^{\ast} &
\frac{73}{15}^{\ast} 
\end{array}
\end{equation}
By using `collared' tiles (which refers to the `border-forcing'
property of \cite{kel}), one can in principle derive all frequencies
in $\D$, see \cite{fw}.  In fact, the frequency of pairs of points
with distance $r=1$ in the table above was calculated this
way. However, the computation of each single frequency requires a
considerable amount of work, and a closed formula for all frequencies
seems out of reach.

\section{Diffraction}

The diffraction of a crystal which is supported on a point lattice in
$\R^d$ is obtained by the Poisson summation formula for Dirac combs
\cite{cor1,cor2}. If $\gG$ is a lattice, the autocorrelation
of the lattice Dirac comb $\delta_{\gG}$ is $\text{dens}(\gG)\,\delta_{\gG}$, 
and the diffraction measure reads 
\begin{equation} \label{eq:psf}
\widehat{\delta}_{\gG} \; =\;  \bigl( \text{dens}(\gG) \bigr)^2 \cdot 
\delta_{\gG^{\ast}}, 
\end{equation}
where $\gG^{\ast}$ is the dual lattice of $\gG$. 

A radial analogue of Eq.~\eqref{eq:psf} is derived in \cite{bfg}. It
is an analogue of the Hardy-Landau-Voronoi formula \cite{ik} in terms
of tempered distributions. Let us first explain this for the example
of the square lattice $\Z^{2}$.  Let $\D^{}_{\smc}$ be the distance
set of $\Z^{2}$, and $\eta^{}_{\smc}(r) := \text{card} \{ x \in \Z^2
\mid |x| = r \}$ the \emph{shelling numbers} of $\Z^2$, see \cite{bg}
for details. Then
\begin{equation} \label{eq:radpsf}
\Bigl( \sum_{ r \in \D^{}_{\smc}}  \eta^{}_{\smc}(r) \, \mu^{}_r
\Bigr)^{\widehat{}} = \sum_{r \in \D^{}_{\smc} } \eta^{}_{\smc}(r)
\, \widehat{\mu}^{}_r = \sum_{r \in \D^{}_{\smc} } \eta^{}_{\smc}(r)
\, \mu^{}_r, 
\end{equation}     
with $\mu_r$ as above. Again, the sum is to be understood as a vague
limit. The fact that $\Z^2$ is self-dual as a lattice (i.e.,
$(\Z^2)^{\ast}=\Z^2$) implies that the same distance set enters all
three sums in \eqref{eq:radpsf}.  In the general case, with an
arbitrary lattice $\gG$, one has to use the distance set of the dual
lattice $\gG^{\ast}$ of $\gG$, in analogy with \eqref{eq:psf}. This
gives the following result \cite{bfg}, valid in Euclidean space of
arbitrary dimension.

\begin{theorem}\label{thm1}
Let $\gG$ be a lattice of full rank in $\R^d$, with dual lattice
$\gG^{\ast}$. If the sets of radii for non-empty shells are
$\D^{}_{\gG}$ and $\D^{}_{\gG^{\ast}}$, with shelling numbers
$\eta^{}_{\gG}(r)=\mathrm{card}\, \{x\in\gG\mid \lvert x\rvert =r\}$ and
$\eta^{}_{\gG^{\ast}}(r)$ defined analogously, the classical Poisson
summation formula has the radial analogue
\begin{equation} \label{eq:rpsf}
\Bigl( \sum_{ r \in \D^{}_{\gG}}  \eta^{}_{\gG}(r) \, \mu^{}_r
\Bigr)^{\widehat{}}
\; = \; \mathrm{dens}(\gG) \sum_{r \in  \D^{}_{\gG^{\ast}} }
\eta^{}_{\gG^{\ast}}(r) \, \mu^{}_r \, ,
\end{equation}
where $\mu^{}_r$ denotes the uniform probability measure on the sphere
of radius $r$ around the origin.
\end{theorem}

Eq.~\eqref{eq:rpsf} also implies
\begin{equation} \label{eq:rpsf2}
\Bigl( \sum_{ r \in \D^{}_{\gG}}  \eta^{}_{\gG}(r) \, \mu^{}_r
\Bigr)^{\widehat{}}
\; = \; \sum_{r \in \D^{}_{\gG} } \eta^{}_{\gG}(r)
\, \widehat{\mu}^{}_r
\end{equation}
in the sense of tempered distributions. This equation is useful in
numerical calculations of pinwheel diffraction spectra.

\subsection{Pinwheel diffraction}

Let us return to the pinwheel pattern $\gL$. In view of
Eq.~\eqref{eq:etarmur} in connection with Proposition~\ref{distset},
the sum in Eq.~\eqref{eq:etarmur} can be recast into a double sum,
\begin{equation} \label{eq:rindsum}
\gamma 
\; = \; \sum_{\ell=0}^{\infty}\;\sum_{r \in 5^{-\ell/2}\D^{}_{\smc}} 
\eta^{}_{\ell} (r) \,\mu^{}_r \, ,
\end{equation}
where the choice of the $\eta^{}_{\ell}(r)$ is not unique, but
restricted by the condition $\sum_{\ell=0}^{\infty}\eta^{}_{\ell} (r)
= \eta(r)$.  If the $\eta^{}_{\ell}(r)$, for fixed $\ell$, could be
chosen to be `lattice-like' --- in the sense that they form a sequence
of shelling numbers for some lattice --- we were in the position to
apply Eq.~\eqref{eq:rpsf} to each inner sum in Eq.~\eqref{eq:rindsum}
individually. Observing that 
\begin{equation} \label{eq:goodsum}
{\bigl(5^{-\ell/2}\,\Z^{2}\bigr)}^{\ast}\;  =\; 
5^{\ell/2}\,\Z^{2}
\end{equation}
this would give rise to terms of the form $\sum_{r \in 5^{\ell/2}
\D^{}_{\smc}}\, \eta^{\,\prime}_{\ell}(r)\, \widehat{\mu}_r$.  Due to
the continuity of the Fourier transform on the space of tempered
distributions, this would imply the diffraction of the pinwheel tiling
to be purely singular.

However, things are not that simple in the case of the pinwheel
tiling. In particular, the $\eta^{}_{\ell}(r)$ cannot be chosen to be
lattice-like, as a consequence of the Delone nature of $\gL$.
Nevertheless, it may still be possible to find coefficients
$\epsilon_{\ell}$, not necessarily positive, such that the autocorrelation
of $\gL$ can be written as
\[ 
\gamma \;=\; \sum_{\ell \ge 0} \epsilon_{\ell}^{} \sum_{r \in 5^{-\ell/2} 
  \D^{}_{\smc}} \eta^{}_{\ell}(r) \, \mu_r\, . 
\]
This general form permits continuous parts in the diffraction
different from the ones arising from \eqref{eq:goodsum}: There may be
singular continuous parts apart from $\{ r\Sb^1 \mid r \in
\D\setminus\{0\} \}$, and even absolutely continuous parts. In fact,
numerical computations indicate \cite{bfg} the presence of an
absolutely continuous part in the diffraction.

Independent of an affirmative answer of the open questions, there is a
striking resemblance with the powder diffraction image of the square
lattice. This is a consequence of Proposition~\ref{distset}. So let us
close this article with a simplified approach to powder diffraction
patterns, and a comparison of the two images.

\begin{figure} 
\begin{center}
\epsfig{width=\textwidth,file=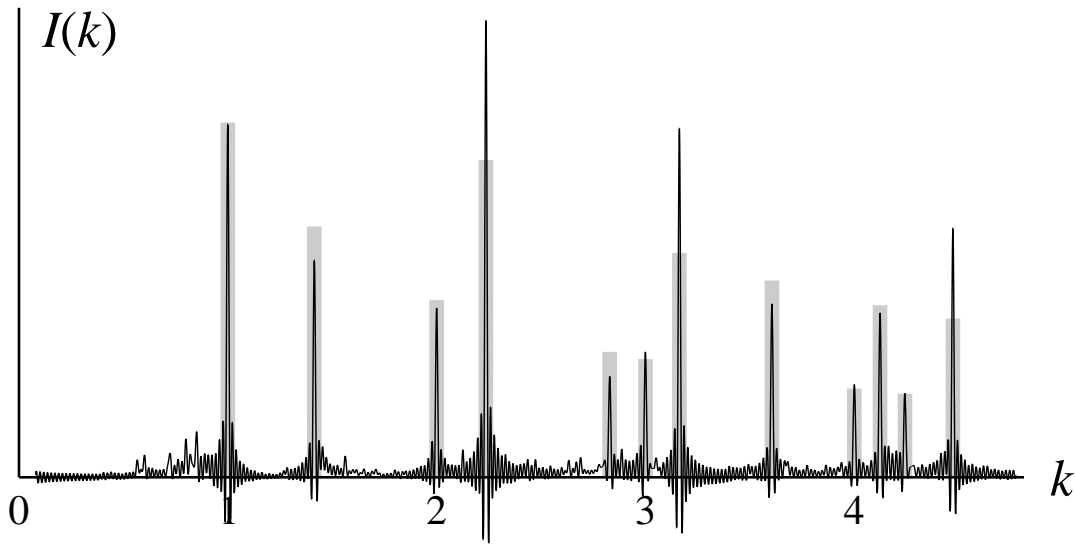}
\end{center}
\caption{Numerical approximation to the radial intensity structure
$I(k)$ of the pinwheel diffraction (solid line) in comparison with the
powder diffraction structure of the square lattice (grey bars). The
relative scale has been chosen such that the heights of the first
peaks at $k=1$ match. Note that the central intensity is suppressed.
\label{fig:pinsq}}
\end{figure}

\subsection{Square lattice powder diffraction}

Instead of performing a diffraction experiment with a large single
crystal, it is often easier to use a probe that comprises many grains
in --- ideally --- random and mutually uncorrelated orientations
\cite{war}. This setting can be modelled mathematically as
follows. Let $R$ be a rotation of infinite order, i.e.,
$R^n\ne\mathrm{id}$ for all integer $n \ne 0$. It follows from Weyl's
lemma that $\{R^n x \mid n \in \N\}$ is uniformly distributed on the
unit circle $\Sb^1$ for any $x \in \Sb^1$. For simplicity, we require
$R\Z^2 \cap \Z^2 = \{0\}$. Then, for large $N$, the set
\[ 
\omega^{}_N = \frac{1}{N} \bigcup_{j=1}^N R^j \Z^2
\] 
can serve as an idealised powder emerging from a two-dimensional
crystal supported on $\Z^2$. Note that the prefactor $1/N$ appears
because we idealise the arrangement of disoriented grains as an
overlay of mutually rotated infinite copies of $\Z^{2}$.  

It is not hard to show \cite{bfg} that the corresponding
autocorrelation is given by
\[ 
\gamma^{}_{\omega_N} =  \frac{N-1}{N}\, \lambda  +\frac{1}{N}\left(\frac{1}{N}
\sum_{j=0}^{N-1}  \delta^{}_{R^j \Z^2}\right), 
\]
where $\lambda$ denotes Lebesgue measure in $\R^2$. The limit of the
bracketed term, as $N \to \infty$, shows perfect circular symmetry,
and is a reasonable approximation of the powder autocorrelation. An
application of Theorem~\ref{thm1} or Eq.~\eqref{eq:radpsf}, yields
\begin{equation} \label{eq:powdiff}
\lim_{N\to\infty}\Bigl( \frac{1}{N}
\sum_{j=0}^{N-1}  \delta^{}_{R^j \Z^2}\Bigr)^{\widehat{}}
\; = \;  \Bigl(\sum_{r \in \D^{}_{\smc}} \eta^{}_{\smc}(r)\, 
\mu_r \Bigr)^{\widehat{}} \; = \;
 \sum_{r \in \D^{}_{\smc}} \eta^{}_{\smc}(r)\, \mu_r \, .
\end{equation} 
This shows that, for an ideal square lattice powder as above, one can
expect a diffraction image that, beyond the central intensity,
consists of concentric rings of radius $r\in\D^{}_{\smc}$ with total
intensity $\eta^{}_{\smc}(r)$. In this simplified version, there is no
absolutely continuous part, though this would be present in a more
realistic model.

To compare the powder diffraction of $\Z^{2}$ with the pinwheel
diffraction, we note that it is sufficient to display the radial
structure. Furthermore, we cannot compare the central intensity,
wherefore we suppress it in both cases. While Eq.~\eqref{eq:powdiff}
gives a closed formula for the intensity of the rings in the powder
diffraction, we currently only have a numerical approximation to the
pinwheel diffraction, based on Eq.~\eqref{eq:rpsf2} and a large patch,
see \cite{bfg} for details. Figure~\ref{fig:pinsq}, which speaks for
itself, shows striking similarities of the singular parts. It is thus
plausible that further investigations in this direction might
ultimately reveal the full nature of the pinwheel diffraction.

\section*{Acknowledgements}

It is our pleasure to thank R.\thinspace V.~Moody and M.~Whittaker for
cooperation and helpful comments. This work was supported by the
German Research Council (DFG), within the CRC 701. UG gratefully
acknowledges conference travel support by The Royal Society.

\end{document}